# Investigating the Perceived Precision and validity of a Field-Deployable Machine Learning-based Tool to Detect Post-Traumatic Stress Disorder (PTSD) Hyperarousal Events

Mahnoosh Sadeghi [1], Farzan Sasangohar [1], Anthony McDonald [1]

1. Department of Industrial and /systems Engineering, Texas A&M University; m7979@tamu.edu; mcdonald@tamu.edu; sasangohar@tamu.edu

* sasangohar@tamu.edu

## Introduction

Post-Traumatic Stress Disorder (PTSD) is a psychiatric condition experienced by individuals after exposure to life-threatening events such as combat exposure, physical assault, and sexual abuse (Kessler et al., 2005). PTSD is a major public health concern and one of the most prevalent mental health disorders in the United States;  over 70% of the U.S. population will experience a traumatic event in their lifetime, and 20% of those affected will go on to develop PTSD, which translates into more than 13 million Americans suffering from this condition at any given time (Sidran Institute, 2018). PTSD is even more common among combat veterans (Kilpatrick et al., 2013) with a recent study suggesting over 24% prevalence (Stefanovics et al., 2020).

PTSD leads to a significant increase in the utilization of healthcare services, especially among combat veterans. The cost of caring for combat veterans usually peaks 30-40 years after a major conflict. For the Iraq/Afghanistan conflicts, it is estimated that cost of veterans' care will peak around 2035 at an estimated $60 billion each year (Geiling et al., 2012). The costs are also expected to increase due to comorbidities including depression, substance abuse, smoking, heart disease, obesity, diabetes, chronic fatigue, and increased dementia.

Major symptoms of PTSD include avoidance, hyperarousal, and reexperiencing the trauma (American Psychiatric Association, 2015). Avoidance symptoms include avoiding activities or cognitions associated with the traumatic event, decreased interest in daily life, and an overall feeling of detachment from one's surroundings. Hyperarousal symptoms include hypervigilance, feelings of irritability, and an exaggerated startle response following a startling event. Other symptoms include anxiety, insomnia, fatigue, anger, and aggression (Carlson et al., 2011). The Diagnostic and Statistical Manual of Mental Disorders (DSM) (American Psychiatric Association, 2015) further divides re-experiencing symptoms into intrusive recollections, recurrent dreams, and flashbacks.

Traditionally, PTSD has been diagnosed using self-reported tools, such as surveys and health questionnaires (e.g., PTSD Checklist for DSM-5 or PCL-5) (American Psychiatric Association, 2013). However, self-reported measures fail to capture isolated and mild cases, and most importantly are not suitable for monitoring or detecting the onset of symptoms (e.g., hyperarousal). Monitoring PTSD hyperarousal events is particularly important because patients may experience intense and severe hyperarousal episodes outside a clinical facility or therapy session. In addition, a major barrier in providing PTSD care is the uncertainty associated with self-management and adherence to therapeutics or medication routines (Rodrigues-Paras et al., 2017). Therefore, there is a vital need to develop effective monitoring systems that provide real-time PTSD hyperarousal detection capability and facilitate data-driven care.

Prior work has shown that several physiological measures including heart rate, heart rate variability, blood pressure, respiratory rate and skin conductance may be correlated with PTSD symptoms (Zoladz & Diamond, 2013). Among these variables, heart rate has shown promise as a reliable PTSD correlate (McDonald et al., 2019; Sadeghi, Sasangohar, & McDonald, 2020; Sadeghi et al., 2019)f and a few studies (Galatzer-Levy et al., 2014, 2017; Saxe et al., 2017) have focused on understanding specific relationships between heart rate and PTSD . However, to our knowledge, only one study has focused on investigating heart rate patterns associated with hyperarousal events (Sadeghi, Sasangohar, Hegde, et al., 2020, 2021; Sadeghi, Sasangohar, McDonald, et al., 2021). That study found that heart rate shows specific patterns during hyperarousal events in terms of fluctuation, stationarity and autocorrelation that is distinct from a typical healthy heart rate pattern in the resting position. Similarly, studies (Galatzer-Levy et al., 2014; Leightley et al., 2019; Liu & Salinas, 2017) have attempted to detect or predict early indicators of PTSD using machine learning algorithms. These studies mostly focused on predicting chances of developing PTSD after a traumatic event. For instance, in one study (Galatzer-Levy et al., 2014) researchers tried to forecast chronic PTSD in individuals based on their early symptoms within 10 days of a traumatic incident. However, all of these studies focused on forecasting and predicting PTSD, and only two studies (McDonald et al., 2019; Sadeghi, McDonald, et al., 2021) focused on PTSD hyperarousal detection based on time and frequency domain features of heart rate and built a machine learning tool for real-time detection of such symptoms.

Despite the promise shown by the above-mentioned machine-learning-based detection tools for PTSD, the validation approaches used in previous research largely relies on theoretical and computational validation methods (e.g., train/test split, cross validation) rather than naturalistic evaluation that accounts for users' perceived precision and validity. Previous research has shown that users' perceptions of physiological changes may not always align well with automated detection of such variables (Zahed et al., 2020) and such misalignment may lead to distrust in automated detection (Fowler, 2021) which may affect adoption or sustainable usage of such technologies. Therefore, the goal of this article is to investigate the perceived precision of the PTSD hyperarousal detection tool developed by Sadeghi et al., 2021 in a home study with a group of PTSD patients. Naturalistic evaluation of such data-driven algorithms may provide foundational insight into efficacy of such tools for non-intrusive and cost-efficient remote monitoring of PTSD symptoms, and will pave the way for their future adoption and sustainable use.

## Methods

To evaluate the efficacy of heart-rate-based machine learning tools to monitor PTSD and assess the perceived precision of one such tool, the PTSD hyperarousal detection algorithm developed by (Sadeghi, McDonald, et al., 2021) was integrated in an application designed for iOS smartphones and smartwatches. In what follows, we summarize the machine learning tool, the integration process, and details of a home study conducted for naturalistic validation of the tool.

**Machine learning algorithm**

The machine learning tool documented in Sadeghi et al. (2021) used the XGBoost algorithm to detect irregular heart rate patterns associated with hyperarousal events (see Sadeghi et al., 2021 for more details). This algorithm was trained based self-reported hyperarousal events of 99 combat veterans who were diagnosed with PTSD. The algorithm had an overall accuracy of 85% on a held aside test set. In addition to heart rate features reported in (Sadeghi, Sasangohar, & McDonald, 2020; Sadeghi et al., 2019), this algorithm used body acceleration to reduce noise in the classification and distinguish between heart rate elevations related to stress and heart rate elevations related to physical activity.

**Integration into a wearable device**

The XGBoost algorithm was integrated in an iOS application (app) for iPhones and iWatches which was previously designed and developed by the Applied Cognitive Ergonomics Lab (ACE-lab) at Texas A&M University (Rao & Sasangohar, 2021). The app was designed exclusively for PTSD self-management. To integrate the machine learning algorithm into the app, we used CoreMLtools function, a framework used for integrating machine learning predictive tools into iOS devices (Marques, 2020; Sahin, 2021), from CoreML package in Python which allows deployment of trained machine learning models on iOS devices. The advantage of CoreML compared to similar frameworks such as TensorFlowlite is its ability to optimize memory usage, and conserve battery life (Tran, 2019).

  The pretrained algorithm integrated in the app could identify irregular patterns in heart rate associated with PTSD hyperarousal events based on real time data collected through iWatch sensors (e.g., heart rate and accelerometer). Upon detection of a hyperarousal event, the tool triggered a notification on the watch interface that asked the user if they perceived any hyperarousal events (Figure 1, top). The users were asked to respond "Yes" or "No" to confirm whether they felt a hyperarousal event or not. The app also had the functionality of self-reporting hyperarousal events by tapping on a specific icon (bell-shaped icon) on the watch face and then confirming that the event happened (Figure 1, bottom). Users could also check their heart rate in real-time by tapping on the heart-shaped icon on the watch interface. Users had to have the application running on their watch at all times, and the app had to be continuously on the watch face to be able to send notifications and collect their data. The app could not be used simultaneously with other activity tracking applications.

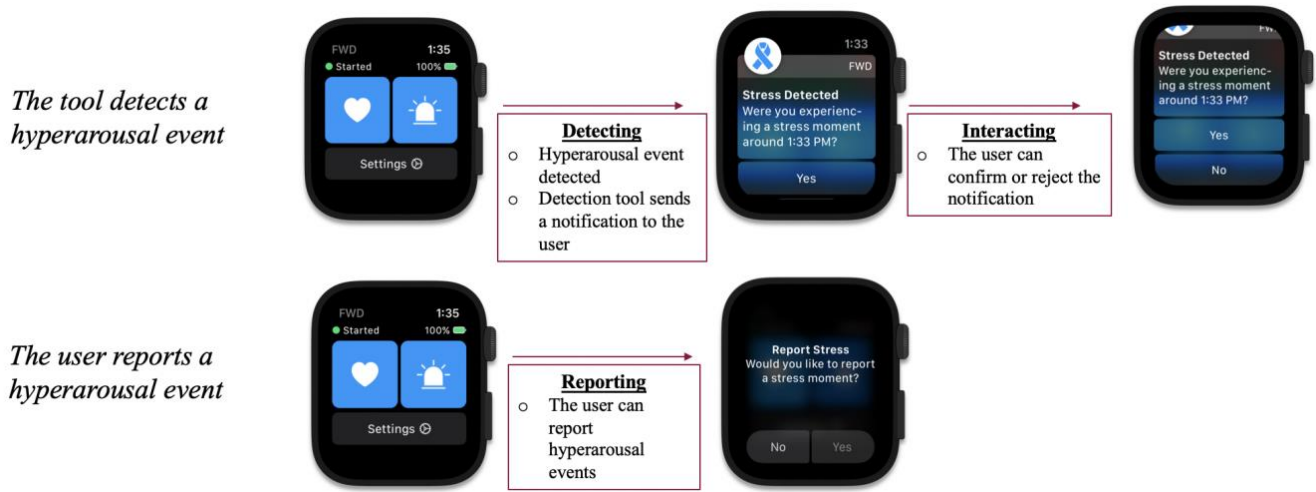

*Figure 1: Illustration of the detection and self-reporting interfaces on iWatch*

**Study process**

*Participants*

Twelve participants were recruited from the Texas A&M University students, staff and faculty population to participate in the study through campus bulk mail. The mean age of all participants was 28 years old (SD = 11.37, range = 18-57). Out of 12 participants, 10 were female and 2 were male. Participants were required to be clinically-diagnosed with PTSD, be over 18 years old, and already own an iPhone (6S or newer) and an Apple iWatch (2nd Gen or newer). Participants were provided with a demographic questionnaire, consent forms, and an Anxiety and Depression Association of America (ADAA) form (cf. American Psychiatric Association, 2015; *Screening for Posttraumatic Stress Disorder (PTSD) | Anxiety and Depression Association of America, ADAA*, n.d.) to prescreen their PTSD diagnosis. In the demographic questionnaire, participants answered questions about their age, gender and PTSD diagnosis. The study was approved by the International Review Board at Texas A&M University (IRB2020-0955DCR).

*Procedure*

Virtual orientation in one-on-one sessions were scheduled with each participant separately. During sessions participants were given instructions on how to install the app on their phone and their watch. Further, participants were instructed on how to interact with the app and the pop-up notifications. Participants were asked to wear the watch continuously for 21 days (between December 20$^{th}$ 2020 – January 25$^{th}$ 2021) and respond to notifications except for when they wanted to charge their devices. They were asked to self-report any instances of hyperarousal. We also emailed participants detailed information about the app installation process and its functionalities. Participants were provided with a list of local licensed therapists to contact in case of an emergency.

*Data collection*

For each participant, we collected data including detected events, participants' responses to symptom detection events (Yes or No), self-reported symptom onset, continuous heart rate data and body acceleration data in three axes: X, Y, and Z. To facilitate and enable data monitoring, we synced the developed the app with Amazon Web Services (AWS). The app stored and uploaded the collected data automatically on AWS at the end of each day of data collection when the user's phone was connected to Wi-Fi. We checked the data on daily basis for each participant to ensure that they are using the app and the detection tool consistently.

**Quantitative Analysis**

Perceived precision was measured by calculating the ratio of correctly detected events (as reported by the user) or true positives to the total number of automated detected events. The equation for perceived accuracy is:

Perceived precision = Number of True Positives / (Number of True Positives+ Number of False Positives)

Where number of true positives is when the tool detected an event and the user responded "Yes" to the notification, and false positive is when the tool detected an event and the user responded "No" to the notification. To investigate time-series trends in the perceived precision , a MannKendall trend test was applied to evaluate the monotonic vs. significant increasing/decreasing trends in the time-series data. For this analysis, we used the Kendall library version 2.2 in RStudio version 3.5.1.

**Interviews and qualitative analysis**

At the end of study, we conducted exit interviews with each participant. The interviews were semi-structured with a focus on user's experience with the detection tool. The participants were asked questions about the accuracy of the tool, their willingness to use this tool, their trust in the detection capability, any issues or barriers related to interaction with the tool, and their expectations from the tool for monitoring of PTSD symptoms. All interviews were virtual through Zoom and were recorded. Additionally, notes were taken during each interview and were later checked with the recorded videos for accuracy. We first transcribed the recordings. We then found a number of themes in the transcriptions and categorized that we will discuss in the results section. Table 1 (in Appendix section) shows all the interview questions.

## Results

### Quantitative results

During 21 days of the data collection 1,244 ($M = 114.33$, $SD = 94.68$) hyperarousal events were detected and 128 ($M = 10.67$, $SD = 10.86$) were reported. Out of the 1,244 detected events, users responded "Yes" 788 ($M = 65.66$, $SD = 74.02$) times to the pop-up notifications indicating a true positive event, and 456 ($M = 38.01$, $SD = 38.55$) times participants reported "No" indicating a false alarm. Figure 2 shows the true positives, false positives, and self-reported events for each participant. Figure 3 shows the distribution for detected true positive and false positive events.

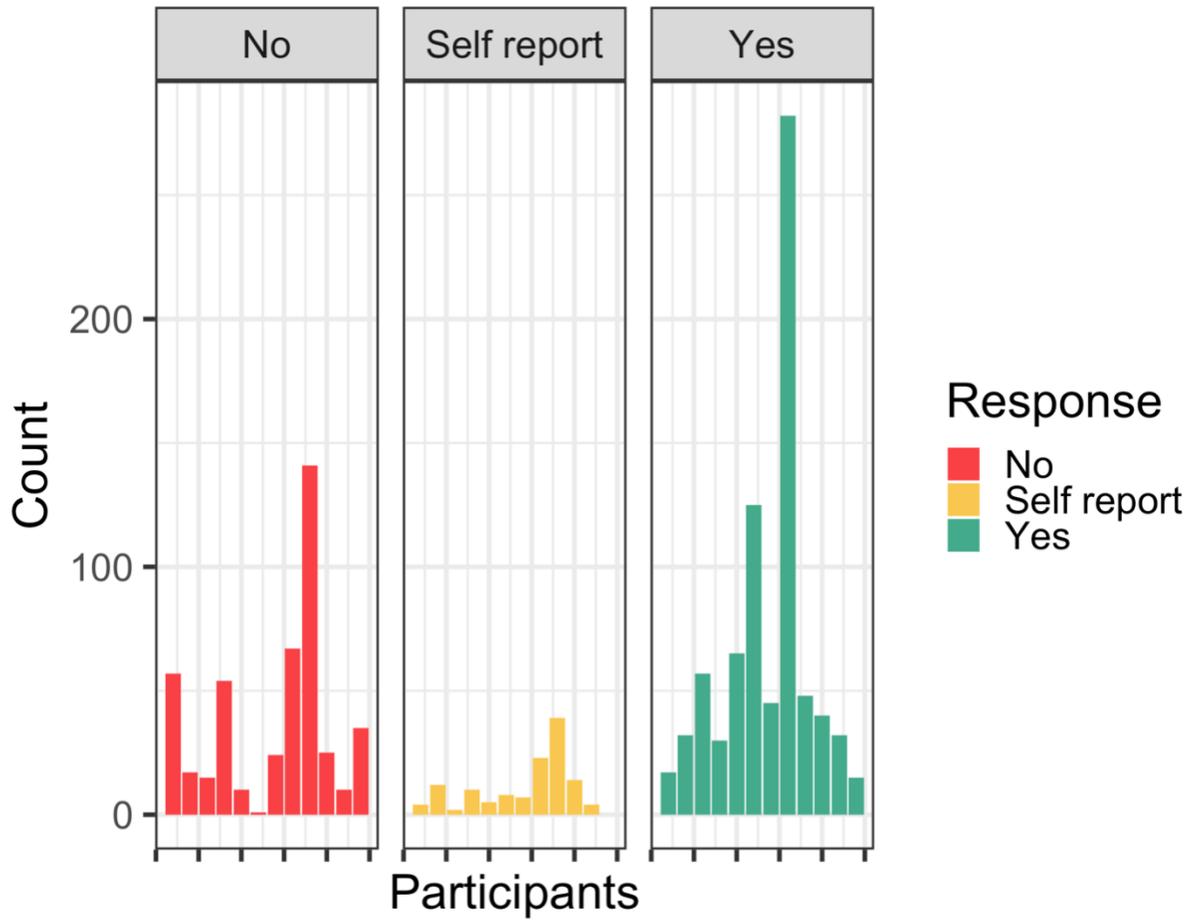

*Figure 2: Count of Yes, No, and Self-Reported events for each participant.*

Based on the analysis, the median perceived precision for all participants was 65.27% (*SD* = 25.9%) ranging from 22.9% to 99.1%. Figure 3 shows the perceived precision for each participant.

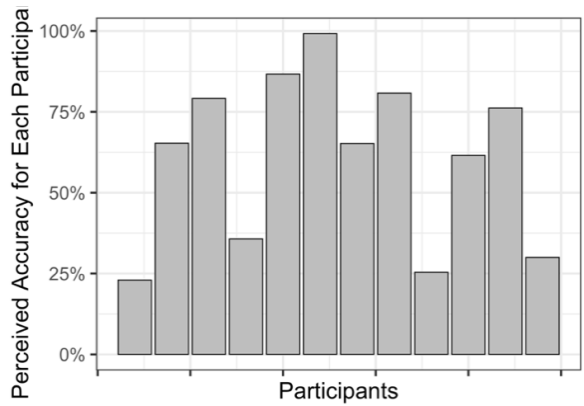

*Figure 3: Perceived precision for each participant*

Further, we investigated heart rate during recorded events. Our findings show that heart rate averaged over a 20 second window (10 seconds before and 10 seconds after the detected events) ranged from 60-182 ($M = 82.04$, $SD = 21.49$) with a median of 76. Figure 4 shows the probability distribution for heart rate during hyperarousal events.

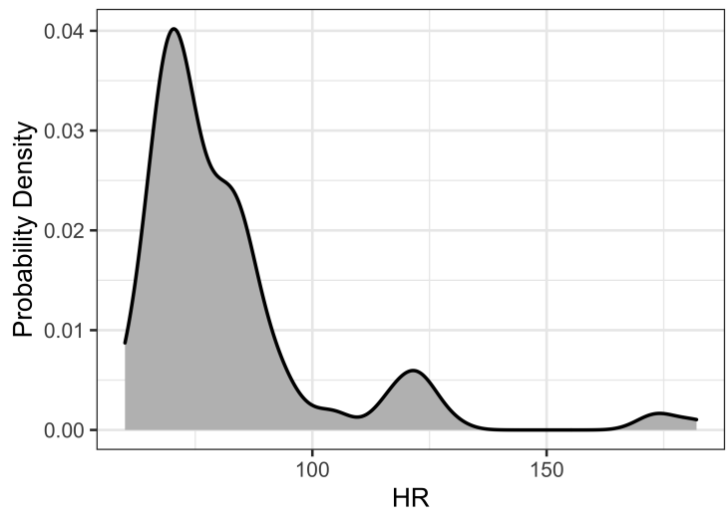

*Figure 4: Heart rate distributions during detected events*

*Trend analysis*

To investigate perceived precision trends, we conducted Mann-Kendall trend analysis on daily perceived precision. The results showed a significant increasing trend for the perceived precision for 11 out of 12 participants (tau = 0.735, *p* < .001). The only exception was a participant who had over 95% perceived precision and their perceived precision did not alter and was uniform throughout the study. We did not observe the same trend for the number of true positives or the number of false positives. Figure 5 shows average perceived precision trend for all participant during 21 days of the study.

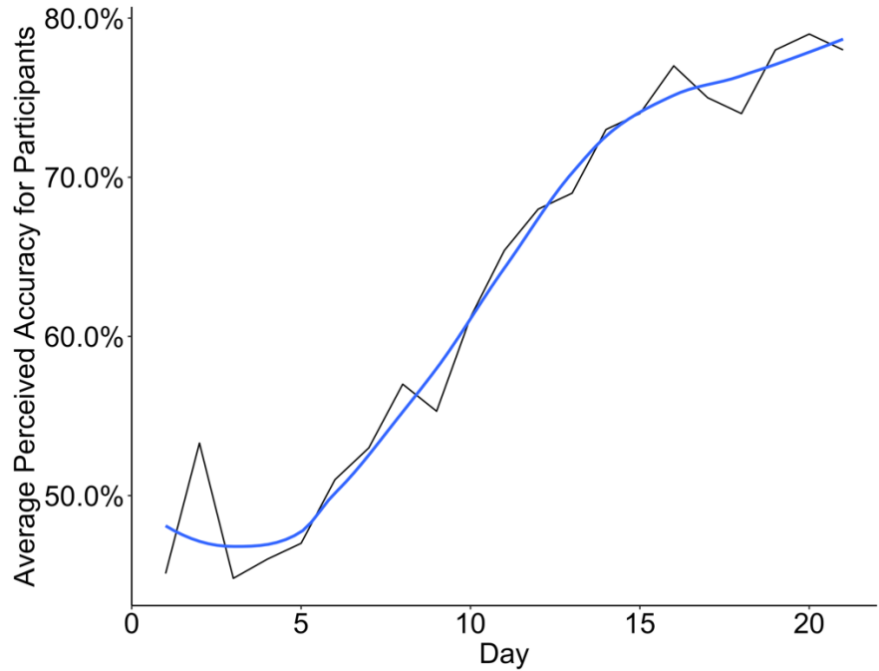

*Figure 5: Average perceived precision trend for all participant during 21 days of the study.*

**Qualitative results**

During exit interviews participants were asked questions about their experience with the tool including their trust in the detection capability, their willingness to use the tool, their expectations from the tool, and main issues with the tool.

*Overall experience*

All participants (12/12) mentioned that they had a positive experience with the detection tool and the app. Participants further commented that the app was easy to use and intuitive.

> *"It was pretty good. It was simple to use, didn't have any problem with it. I enjoyed being able to go back and look back at my data." - Participant 9.*

The majority of participants (10/12) mentioned that it was helpful for them to be aware and mindful of their hyperarousal events.

> *"I liked that it kind of pinged me to try to think about it. It was right most of the times. It provides another opportunity for you to stop and reflect, but again, depending on how deeply in the therapy that person is or how aware that person is, the app can do so much. The rest of it is on the use to sit and think about it." Participant 2.*

> *"I thought that it was helpful, especially one of the things that I'm working on in therapy right now is actually gauging my stress and anxiety level and how to cope appropriately, so I thought that it was definitely helpful"* Participant 9.

> *"Yes, I think it was helpful in the fact that it made me more aware of what was going on, and I was able to recognize it, be aware of the moment, and bring myself back down, or maybe even figure out what was causing stress and what not."* Participant 10.

However, 2 participants explained that being reminded of hyperarousal events has a negative reinforcement effect.

> *"I think it made me hypersensitive of it, so I would notice it more usually than I would."* Participant 1.

> *"For me because the way I experience PTSD is to shut down, I think it would be helpful maybe to just receive a report at the end of the day that says hey these are all the times today that we noticed hyperarousal instead of being notified in the moment where it might amplify it more than it should"* Participant 3.

*Perceived precision*

Most participants (11/12) mentioned that the tool could accurately detect their hyperarousal events. These participants reported that they trust the tool's detection capability for hyperarousal events.

> *"Yes definitely! I am personally grounded in my routines, just one day my girlfriend and I were getting dinner. She forced me to try something new and the app immediately asked if I was [hyperaroused]."* Participant 7.

> *"I think it's relatively trustworthy, there were a couple of times that it detected that weren't accurate but other than that it was incredibly consistent."* Participant 9.

In particular, participants perception of percentage of false alarms ranged from 0-50% with an average of 18.58%. This number is roughly 16% lower than the reported false alarm rates from the self-reports (34.73%) (i.e., percentage of No responses to the detection notification). Three participants (3/12) reported that they did not receive any false alarms, one of whom had an objective perceived precision of almost 100% (i.e., responded YES to all detection notification). Despite such perception, the data from these three participants showed up to 10% reported false alarms. All participants (12/12) found the frequency of false alarms to be acceptable. Participants commented that notifications were easy to respond to and not interruptive.

> *"I would say like it was just a quick Yes or No, it was not interruptive because it was so quick, and I just had to look at my watch to say Yes or No" Participant 11.*

> *"Yeah, it is acceptable. I think that with any stress detection tool there is going to be certain levels of inaccuracy that you have to understand and account for. This was a low-enough number." Participant 9.*

Some participants (3/12) mentioned that the app sometimes falsely notified them when they were engaging to an activity; while others mentioned that they received false alarms when they were in relaxed position.

> *"I think most of the times it was when I was sitting down, and I was like probably in a conversation with somebody, or I was trying to figure out a work project, or just thinking about my day. I don't think that I was doing any sort of active thing" Participant 7.*

> *"I was usually engaged in an activity doing something like cleaning out the garage, you know, that was moving around. And most of the other times was frustration with work." Participant 8.*

However, one participant mentioned difficulties in assessing the accuracy of detected events due to issues related to lack of self-awareness.

*"I don't know! It requires the person to be self-aware enough to know that they are stressed. I have to sit in and check with myself to see if I am having a hyperarousal moment." Participant 2.*

### *Interaction with notifications*

Overall, 3.57% of detection notifications did not receive a response. Over 50% of the participants (7/12) mentioned that they missed notifications because of their surrounding situation such as discreetness or the time that they received notifications.

*"It's just very hard to respond to something like that, and also just when people are around, sometimes it's just you don't want them to see, you know, because people ask questions about it" Participant 5.*

*"A lot of these notifications came at night, and I was not sure, so I dismissed them" Participant 2.*

### *Expectations for notifications*

Over 50% of the participants (7/12) indicated that they liked the visual notification and the vibration. Participants explained that the notification message was clear, the vibration was sensible, and it was easy to respond to notification.

*"I like how it was Yes or No, just didn't have to think about it that much, it was just simple. I like the vibration because like when my watch detected something I could sense it. I also prefer vibration to -*

> *sound and other things because other things stress me out more."* Participant 4.
>
> *"I liked the message and can't think about a way that I would change it. I like the vibration more I think than something like a sound. Because other people can hear the sound. I had a friend in town and I told him about this app, and it got to the point that when I did have the sound on then he knew that I was stressed and I did not want anybody else to know. And the sound can sometimes be triggering"* Participant 6.

A few participants (4/12) commented that they did not like that the notification message stayed on their watch face since it was noticeable by others.

> *"I liked the vibration, but I didn't like how the alert stayed on the screen, just because everybody could see it."* Participant 1.

A few participants (4/12) elaborated that the vibration was not strong enough and they missed notifications because of that.

> *" I think during the day the amount of vibration is fine, but during the night if you want to catch somebody sleeping it needs to be stronger."* Participant 2.

> *"I think the vibration could have been a little stronger. I am a very active person and I use my hands a lot at work, so I think sometimes the motion outweighs the vibration"* Participant 6.

> *"It was a pretty low vibration sensation and so if I had been like in class or something like that, it probably would have been a lot harder to pick up on. I'd like to get a second notification shortly after if there is no response to the initial."* Participant 9.

***Expectations for post-detection interaction***

Most participants (10/12) expected the tool to support users develop coping skills post-detection. Among the coping skills, breathing exercises (8/12), and meditation such as focusing exercises (5/12) were mentioned by some.

> *"I think telling me to breathe would be good or maybe a message that says you are OK right now, or some type of encouraging message that will bring you back to reality."* Participant 3.

However, a few participants (2/12) mentioned that improving self-awareness through using the detection tool is sufficient for them and the scope of the tool should be limited to awareness rather than being directed to coping activities.

> *"For me it was just enough to acknowledge and like actively press something that said Yes I am feeling the stressful moment instead of just like trying to keep it in, for me that was enough."* Participant 2.

> *"I don't really expect it to do anything. I think it brings the awareness to myself you know that its detecting something. I'll be able to bring myself back down"* Participant 10.

***Acceptance and barriers to adopt***

All participants (12/12) mentioned that they would use the tool on a daily basis with a few adjustments. The main issue that prevented most participants from using the app continuously was the battery life. Continuous monitoring of physiological data through watch sensors affects the battery life significantly. Most participants (10/12) commented on this issue and explained that they had to charge their watch more frequently (e.g., 1 to 2 times a day compared to once every 48 hours).

> *"I really would use it. I liked monitoring my heart rate and I liked it really pointing out the event, cause then I could sit there and almost predict them and then be able to deal with them from there."* Participant 12.

> *"Yeah, I would like to use the app on a daily basis. I guess the only thing is that it takes up a lot of battery power when I'm wearing it. That was the main issue like having to charge it every night."* Participant 4.

> *"I would definitely use it, I think it would be helpful. The downside was the battery life. It would be nice if it did not drain the battery so you could wear it for a longer period of time."* Participant 8.

## Discussion

This work describes a novel mixed-methods analysis for naturalistic validation of a machine learning algorithm to detect PTSD hyperarousal events. While PTSD treatments and psychophysiological assessments in the laboratory setting have been well documented, there is a need for additional research to bridge the gap in continuous monitoring of PTSD symptoms to improve self-management and inform therapeutic care. Although naturalistic studies have been conducted to develop machine learning tools to detect PTSD hyperarousal events (e.g., McDonald et al., 2019), to our knowledge, this study is the first to evaluate the perceived precision of real-time PTSD hyperarousal detection in naturalistic settings.

Our findings suggest that the average perceived precision of the detected events was about 65%, indicating that majority of participants agreed with the automated real-time detection of an irregular heart rate pattern associated with PTSD hyperarousal. The range of the perceived precision showed substantial between participants' variability in perception of accuracy (Range: 23%-99%) with false alarm rates as high as 77%. However, our qualitative suggest that false alarms were tolerated well and accepted by the users.

These findings highlight the importance of naturalistic evaluation and validation of machine learning tools. While common approaches in machine learning accuracy measurements and performance evaluation may provide initial evidence of objective efficacy, in applications involving interactions with humans, subjective perceptions may be exhibit misalignment with empirical evidence. This misalignment has been shown in various comparisons of objective and subjective evidence (Chellappa et al., 2018; Kosmadopoulos et al., 2017). In particular, in our study, we noticed a significant disparity between the theoretical precision for the developed machine learning algorithm and the users' perceived precision (70% vs. 65% for the algorithm used in this study, respectively). Therefore, sole reliance on theoretically-driven performance metrics without accounting for users' perception of accuracy may result in negative impacts on sustainable usage and acceptance of machine-learning-based tools.

Another interesting finding was that perceived precision increased over time for majority of participants. Given the known correlation between perceived precision of automation and trust (Findley, 2015; Merritt, 2011), this finding may suggest that while users may have initially distrusted the tool, longitudinal exposure to true positives increased and calibrated their trust in the tool's capability in detecting hyperarousal events. Such initial mistrust of automation has been shown in other research (Fowler, 2021; Tenhundfeld et al., 2019). This is also supported by findings from the exit interviews. Several participants elaborated on the process of trust-building over time. For example, participant 12 explained that: *"the first week I was denying notifications because I didn't think they were stressful enough to be considered stressful events, and then I realize all of them probably were stressful events, but I think I mentally have a threshold of what is considered a hyperarousal event. In that thought, I don't think there were any real false positives"*. This important finding highlights the importance of conducting naturalistic validation studies longitudinally. While we did not notice a decline in trust or usage over time, future work may use longer study durations to understand users' behavior over an extended period. Future improvements in personalized and reinforcement learning (i.e., adjusting the detection based on participants' responses) may also improve the calibration in trust and needs to be examined.

      The results provided preliminary evidence that the tool detected hyperarousal events, albeit with a high false alarm rate. However, overall, the tool was received well, and most participants found it helpful in increasing self-awareness and being mindful of hyperarousal events. This is in line with previous research indicating that stress awareness facilitates and improves stress

management (Morris et al., 2010). However, a few participants elaborated on the negative reinforcement of the detection tool's notifications to trigger further stress. It is well documented that alarms and notifications may result in a startle effect or further stress (Taylor et al., 1996). Additionally, false alarms might trigger panic attacks and put the body in a fight-or-flight episode which may exacerbate anxiety and stress (Fowles, 2019). While this study focused on subjective validation and perceived precision of detection, more work is warranted to investigate user-centered design of notifications and alerts and other interactive features to identify and mitigate stress-inducing design factors and improve users' experience. In addition, while majority of participants in this study valued real-time notifications, a few preferred on-demand access to such information. This may suggest the importance of personalized notification settings to tailor interventions to variety of users.

There are several limitations that impose constrains on the generalizability of the findings reported in this work. First, the sample size was small, limiting the ability to conduct inferential statistics. We tried to partly address this limitation by conducting a longitudinal study over 21 days. However, more work is needed to confirm our findings with a larger sample size. Second, we did not account for the PTSD severity or other co-morbidities. Third, the analysis of perceived precision assumes a collectively-exhaustive account of all perceived hyperarousal events.

While we emphasized the importance of self-reporting to our participants, events may have been under-reported. Finally, while this study addresses an important gap in naturalistic evaluation of machine-learning-based tools for detecting PTSD hyperarousal events (and broadly stated "stress"), there are several limitations and barriers associated with conducting a home study using wearables. For example, real-time monitoring required significant computational power which had significant impact on the smartwatch battery life. This barrier resulted in but also potential impact on participants' self-reporting behavior and overall impressions of the tool. More work is warranted to optimize the machine learning tools for computational efficiency and improve the integration into wearables technologies.

## Conclusion

Non-intrusive monitoring tools for PTSD hyperarousal events to improve self-management and self-awareness is a timely need and a general gap in research. This paper contributes to the body of literature by providing preliminary evidence of efficacy for one such tool while highlighting the importance of naturalistic evaluation of machine-learning-based detection tools accounting for users' perceptions and interactions. Future work in utilizing user-centered design methods and just-in-time evaluation techniques will help improving the design of effective PTSD continuous monitoring and self-management tools that address an important gap in current PTSD care models and is expected to have a positive impact on quality of life and PTSD health outcomes.

# References

American Psychiatric Association. (2013). *Diagnostic and statistical manual of mental disorders (DSM-5®)*. American Psychiatric Pub.

American Psychiatric Association. (2015). *Depressive Disorders: DSM-5® Selections*. American Psychiatric Pub.

Carlson, K. F., Kehle, S. M., Meis, L. A., Greer, N., MacDonald, R., Rutks, I., Sayer, N. A., Dobscha, S. K., & Wilt, T. J. (2011). Prevalence, assessment, and treatment of mild traumatic brain injury and posttraumatic stress disorder: A systematic review of the evidence. *The Journal of Head Trauma Rehabilitation*, *26*(2), 103–115.

Chellappa, S. L., Morris, C. J., & Scheer, F. A. (2018). Daily circadian misalignment impairs human cognitive performance task-dependently. *Scientific Reports*, *8*(1), 1–11.

Fowler, M. J. (2021). *Assessing the Development of Operator Trust in Automation* [PhD Thesis].

Fowles, D. C. (2019). Motivational approach to anxiety disorders. In *Anxiety: Recent developments in cognitive, psychophysiological, and health research* (pp. 181–192). Taylor & Francis.

Galatzer-Levy, I. R., Karstoft, K.-I., Statnikov, A., & Shalev, A. Y. (2014). Quantitative forecasting of PTSD from early trauma responses: A machine learning application. *Journal of Psychiatric Research*, *59*, 68–76.

Galatzer-Levy, I. R., Ma, S., Statnikov, A., Yehuda, R., & Shalev, A. Y. (2017). Utilization of machine learning for prediction of post-traumatic stress: A re-examination of cortisol in the prediction and pathways to non-remitting PTSD. *Translational Psychiatry*, *7*(3), e1070–e1070.

Geiling, J., Rosen, J. M., & Edwards, R. D. (2012). Medical costs of war in 2035: Long-term care challenges for veterans of Iraq and Afghanistan. *Military Medicine*, *177*(11), 1235–1244.

Kessler, R. C., Berglund, P., Demler, O., Jin, R., Merikangas, K. R., & Walters, E. E. (2005). Lifetime prevalence and age-of-onset distributions of DSM-IV disorders in the National Comorbidity Survey Replication. *Archives of General Psychiatry*, *62*(6), 593–602.

Kilpatrick, D. G., Resnick, H. S., Milanak, M. E., Miller, M. W., Keyes, K. M., & Friedman, M. J. (2013). National estimates of exposure to traumatic events and PTSD prevalence using DSM-IV and DSM-5 criteria. *Journal of Traumatic Stress*, *26*(5), 537–547.

Kosmadopoulos, A., Sargent, C., Zhou, X., Darwent, D., Matthews, R. W., Dawson, D., & Roach, G. D. (2017). The efficacy of objective and subjective predictors of driving performance during sleep restriction and circadian misalignment. *Accident Analysis & Prevention*, *99*, 445–451.

Leightley, D., Williamson, V., Darby, J., & Fear, N. T. (2019). Identifying probable post-traumatic stress disorder: Applying supervised machine learning to data from a UK military cohort. *Journal of Mental Health*, *28*(1), 34–41.

Liu, N. T., & Salinas, J. (2017). Machine learning for predicting outcomes in trauma. *Shock: Injury, Inflammation, and Sepsis: Laboratory and Clinical Approaches*, *48*(5), 504–510.

Marques, O. (2020). Machine Learning with Core ML. In *Image Processing and Computer Vision in iOS* (pp. 29–40). Springer.

McDonald, A. D., Sasangohar, F., Jatav, A., & Rao, A. H. (2019). Continuous monitoring and detection of post-traumatic stress disorder (PTSD) triggers among veterans: A supervised machine learning approach. *IISE Transactions on Healthcare Systems Engineering*, *9*(3), 201–211.

Morris, M. E., Kathawala, Q., Leen, T. K., Gorenstein, E. E., Guilak, F., DeLeeuw, W., & Labhard, M. (2010). Mobile therapy: Case study evaluations of a cell phone application for emotional self-awareness. *Journal of Medical Internet Research*, *12*(2), e10.

Rao, A. H., & Sasangohar, F. (2021). Designing for Veterans. In *The Patient Factor* (pp. 107–124). CRC Press.

Sadeghi, M., McDonald, A. D., & Sasangohar, F. (2021). Posttraumatic Stress Disorder Hyperarousal Event Detection Using Smartwatch Physiological and Activity Data. *ArXiv Preprint ArXiv:2109.14743*.

Sadeghi, M., Sasangohar, F., Hegde, S., & McDonald, A. (2021). Investigating Heart Rate Patterns During Hyperarousal Events Among Veterans Who Have Posttraumatic Stress Disorder (PTSD). *Proceedings of the International Symposium on Human Factors and Ergonomics in Health Care*, *10*(1), 91–91.

Sadeghi, M., Sasangohar, F., Hegde, S., & McDonald, A. (2020). Understanding Heart Rate Reactions to Posttraumatic Stress Disorder (PTSD) Among Veterans. *Proceedings of the Human Factors and Ergonomics Society Annual Meeting*, *64*(1), 780–780.

Sadeghi, M., Sasangohar, F., & McDonald, A. (2019). Analyzing Heart Rate as a Physiological Indicator of Post-Traumatic Stress Disorder: A Scoping Literature Review. *Proceedings of the Human Factors and Ergonomics Society Annual Meeting*, *63*(1), 1936–1936.

Sadeghi, M., Sasangohar, F., & McDonald, A. D. (2020). Toward a taxonomy for analyzing the heart rate as a physiological indicator of posttraumatic stress disorder: Systematic review and development of a framework. *JMIR Mental Health*, *7*(7), e16654.

Sadeghi, M., Sasangohar, F., McDonald, A. D., & Hegde, S. (2021). Understanding Heart Rate Reactions to Post-Traumatic Stress Disorder (PTSD) Among Veterans: A Naturalistic Study. *Human Factors*, 00187208211034024.

Sahin, Ö. (2021). Introduction to Apple ML Tools. In *Develop Intelligent iOS Apps with Swift* (pp. 17–39). Springer.

Saxe, G. N., Ma, S., Ren, J., & Aliferis, C. (2017). Machine learning methods to predict child posttraumatic stress: A proof of concept study. *BMC Psychiatry*, *17*(1), 1–13.

*Screening for Posttraumatic Stress Disorder (PTSD) | Anxiety and Depression Association of America, ADAA*. (n.d.). Retrieved July 28, 2021, from https://adaa.org/screening-posttraumatic-stress-disorder-ptsd

Sidran Institute. (2018). *Post Traumatic Stress Disorder Fact Sheet*. https://www.sidran.org/wp-content/uploads/2018/11/Post-Traumatic-Stress-Disorder-Fact-Sheet-.pdf

Stefanovics, E. A., Potenza, M. N., & Pietrzak, R. H. (2020). PTSD and obesity in US military veterans: Prevalence, health burden, and suicidality. *Psychiatry Research*, *291*, 113242.

Taylor, S., Woody, S., Koch, W. J., McLean, P. D., & Anderson, K. W. (1996). Suffocation false alarms and efficacy of cognitive behavioral therapy for panic disorder. *Behavior Therapy*, *27*(1), 115–126.

Tenhundfeld, N. L., de Visser, E. J., Haring, K. S., Ries, A. J., Finomore, V. S., & Tossell, C. C. (2019). Calibrating trust in automation through familiarity with the autoparking feature of a Tesla Model X. *Journal of Cognitive Engineering and Decision Making*, *13*(4), 279–294.

Tran, X. T. (2019). *Applying computer vision for detection of diseases in plants* [PhD Thesis]. Iowa State University.

Zahed, K., Sasangohar, F., Mehta, R., Erraguntla, M., & Qaraqe, K. (2020). Diabetes Management Experience and the State of Hypoglycemia: National Online Survey Study. *JMIR Diabetes*, *5*(2), e17890.

Zoladz, P. R., & Diamond, D. M. (2013). Current status on behavioral and biological markers of PTSD: A search for clarity in a conflicting literature. *Neuroscience & Biobehavioral Reviews*, *37*(5), 860–895.

# Appendix

*Table 1: List of exit interview questions*

| Question |
|---|
| 1. How was your overall experience with the app and the detection tool? |
| 2. Overall, do you think the tool can accurately detect hyperarousal events? |
| 3. Do you trust the hyperarousal detection capability of the tool? Why/why not? |
| 4. How many notifications a day did you receive on average? |
| 5. What percentage of these notifications do you think were false alarms? |
| 6. Do you find this level of false alarm to be acceptable? |
| 7. My data shows you responded (either said yes or no) to notifications X percent of the time. Can you describe why you didn't respond to some alerts? [Probe: were you engaged in any sort of activity? What were some of these common activities?] |
| 8. Can you describe what you feel during a hyperarousal event? what happens to you physiologically? Are there any effects on your heart rate? |
| 9. Let's discuss the alert message itself. What type of message do you expect to see on the alert when the tool detects a hyperarousal moment? |
| 10. What would you expect from a tool (for example a mobile app or smartwatch app) to do for you when it detects a hyperarousal event? |
| 11. Overall, was it helpful to be reminded of your hyperarousal events? |
| 12. Would you use this detection tool on a daily basis? What are some of the barriers in using this tool continuously? |
| 13. Do you have any other comments? |